\documentstyle[12pt,epsf]{article}

\begin{document}

\begin{flushright}
IC-95-217\\
UMD-PP-96-14
\end{flushright}

\begin{center}
{\Huge A low $\alpha_s$ and its consequences for unified model building}

\vskip 1.5 cm

{\bf
Biswajoy Brahmachari\footnote{Electronic address:
biswajoy@ictp.trieste.it} \\
\medskip
High Energy Section\\
International Centre For Theoretical Physics\\
34100 Trieste, Italy.\\}

\medskip

and\\

\medskip

{\bf Rabindra N. Mohapatra\footnote{Electronic
address: rmohapatra@umdhep.umd.edu}\\
\medskip
Department of Physics\\
University of Maryland\\
College Park, Maryland 20742, USA.}

\vskip 1.5cm
\end{center}

\begin{abstract}
We review various ways of obtaining consistency between the idea of
supersymmetric grand unification and an apparent low value of $\alpha_s
\sim 0.112$ indicated by several low energy experiments. We argue that to
reconcile the low value of $\alpha_s$ with the predictions of
supersymmetric GUTs, we need to go beyond the standard minimal
supersymmetric GUT scenario and invoke new physics either at
$10^{11}-10^{12}$ GeV, or at the GUT scale.
\end{abstract}

\vskip 0.5cm

\newcommand{\be}{\begin{equation}}
\newcommand{\ee}{\end{equation}}
\newcommand{\bea}{\begin{eqnarray}}
\newcommand{\eea}{\end{eqnarray}}
\def\tl{{\tilde{l}}}
\def\tL{{\tilde{L}}}
\def\bd{{\overline{d}}}
\def\tL{{\tilde{L}}}
\def\a{\alpha}
\def\b{\beta}

\newpage

\section{Introduction}
Recently there has been a lot of interest in the difference in the
measured values of $\alpha_s$ in high and low energy experiments. A
very useful summary of the issues have been given in a recent paper by
Shifman\cite{shifman} suggesting that the observed discrepancy between
the higher values of $\approx 0.125$ for $\alpha_{s}(M_Z)$ derived
from the global fit of LEP/SLC data assuming only the SM particle
content and interactions on the one hand and lower values near
$\approx 0.11$ derived from the low energy data such
as deep inelastic electron scattering \cite{altarelli,vir}, lattice
calculations \cite{el-kadra}
involving the upsilon and the $J/\Psi$ system etc on the other should be
considered to be an indication of the presence of new
physics\cite{langa}. If this new physics is identified with the
supersymmetric version of the standard model at low energy, then one
can attempt to do a global fit to all LEP/SLC data including the
supersymmetric particles and interactions and see if indeed the high
value of $\alpha_{s}(M_Z)$ indicated there is lowered. Such an
analysis has been carried out recently by several groups\cite{kane},
 who have shown that if the stop and the chargino masses
are kept below a 100 GeV, then there are new contributions to the
$Z\rightarrow b\overline{b}$ decay which increase its decay width.  In
the presence of these contributions, the global fit to LEP/SLC data indeed
leads to a value for $\alpha_{s}(M_Z)
\simeq .112$ which is what the low energy data give for this parameter.
Of course it should be noted that the values of parameters used in the
above discussion may not be obtainable in simple SUSY GUT theories.
It should be noted that there are a number of other suggestions that
could also lead to a higher value for the $Z\rightarrow b\overline{b}$
width. It could very well be that one of these scenarios rather than
the SUSY contribution is at the real heart of the problem. But for our
discussion of unification, it important that either the experimental
value of the $Z\rightarrow b\overline{b}$ come down or that the
supersymmetric scenario provide an explanation for its enhancement
over the standard model value so that the value of $\alpha_{s}(M_Z)$
gravitates towards its lower value. As we discuss below, this low value
will have profound implications for the nature of SUSY GUT theories.

The LEP measurements of the gauge couplings
$\alpha_1,\alpha_2$ and $\alpha_{s}$ combined with the
coupling constant evolution dictated by the minimal supersymmetric
standard model, have in the past three years led to speculations that
the present data while providing overwhelming support for the standard
model may in fact be indicating that the next level of physics
consists of a single scale supersymmetric grand unified theory, with
new physics beyond supersymmetry appearing only at the scale of
$10^{16}$ GeV\cite{amaldi}. This has generated a great deal of
excitement and activity in the area of supersymmetric grand unified
theories(SUSY GUT). Consequently, the renormalization group (RG) analysis
have become more
refined over the time including the various low and high scale threshold
corrections. The effect of low energy threshold corrections has been
included in the simple step function approximation as well as by including
a detailed mass dependent renormalization scheme \cite{bagger, mar, chan}.
These analysis conclude that the effect of the low energy threshold effects
is to increase the predicted value of $\a_s $. The GUT scale threshold
corrections to $\a_s$ come mainly from two sources in the minimal scheme-
the doublet-triplet splitting and the mass splitting among the heavy scalars
present at the GUT scale. The threshold effects from the well known
doublet-triplet splitting always increases the
prediction of $\a_s$. The threshold effects from
the splitting among the heavy scalars can be of either sign depending on
the mass spectrum of the heavy scalars. In the minimal SU(5) GUT the heavy
scalars reside in the 24 dimensional adjoint representation of SU(5) and the
spectrum the masses of these heavy scalars are quite constrained. A
combined analysis including the heavy and light threshold corrections in
the minimal SU(5) GUT predicts a value of $\a_s$ around 0.126\cite{bagger}
 when the
superpartner masses are at the TeV scale. If the s-particle masses are
less than 1 TeV the prediction of $\a_s$ increases further. For example,
for s-particle masses around 500 GeV, the predicted value of $\a_s$ is
0.130, even beyond the value measured at LEP. Clearly, this leads to a
conflict. On the one hand to reduce the prediction of $\a_s$ in a SUSY
GUT one needs a high value of the s-particle masses; on the other hand, to fit
the LEP/SLC data with a low $\a_s$ in a supersymmetric model, one needs the
stop and the chargino below 100 GeV.

In this brief review, we will summarize the arguments leading one to
conclude that if the value of
the QCD fine structure constant $\alpha_{s}(M_Z)$ at the weak scale turns out
to be around $0.112$ as is indicated by several low energy experiments then
to reconcile this value with the predictions for supersymmetric GUTs
one necessarily needs new physics beyond the usual minimal SUSY GUT
scenarios, either at $10^{11}-10^{12}$ GeV or at
the GUT scale. Several new physics possibilities
 have been identified in recent
literature such as, (a) introducing non-universality of gaugino masses at
the GUT scale; (b) introducing heavy threshold effects coming from the heavy
scalar multiplets at the GUT scale, when the scalar sector of the minimal
SUSY SU(5) model is altered to include $50$, $\overline{50}$, and $75$
representations of SU(5); (c) introducing a superstring-inspired scalar
spectrum in a supersymmetric SO(10) GUT and making room to incorporate an
intermediate B-L symmetry breaking scale; (d) introducing an extra mini
doublet-triplet splitting near the GUT scale. This paper is organized as
 follows. In section 2, we state the basic notions; in section 3 we
outline how the non-universality of
gaugino masses can lead to a lowering of the prediction of $\a_s$; in
section 4, we describe the heavy threshold corrections in the SU(5) model
leading to the change in the prediction of $\a_s$; in section 5, we
describe the introduction of the string inspired scalar spectrum in the
supersymmetric SO(10) GUT - and the consequent lowering of the prediction
of $\a_s$ and in section 6, we describe the extra mini doublet triplet
splitting in the GUT scale. In section 7, we present our conclusions.

\section{Basic notions}

To illustrate our basic procedures, let us consider a toy example in which
all the superpartners are degenerate at the scale $M_Z$, excepting the
gluinos and the winos, which are somewhat heavier than the scale
$M_Z$. In such a scenario, the three gauge couplings at the scale $M_Z$ can
be related to the unification coupling by the relations,
\begin{eqnarray}
2 \pi \alpha^{-1}_s~(M_Z)&=& 2 \pi \alpha^{-1}_U + b^{susy}_s \ln {M_U
\over M_{\tilde{g}} } + [ b^{susy}_s - \Delta_{\tilde{g}}] \ln
{M_{\tilde{g}} \over M_Z}, \nonumber\\
2 \pi \alpha^{-1}_2~(M_Z)&=& 2 \pi \alpha^{-1}_U + b^{susy}_2 \ln
{M_U \over M_{\tilde{w}} } + [ b^{susy}_2 - \Delta_{\tilde{w}}]
\ln{M_{\tilde{w}} \over M_Z},
\nonumber\\
2 \pi \alpha^{-1}_1~(M_Z)&=& 2 \pi \alpha^{-1}_U + b^{susy}_1 \ln{M_U
\over M_Z}, \label{rge}
\end{eqnarray}
where, $b^{susy}_i$ are the usual one-loop supersymmetric beta function
coefficients and $\Delta_{\tilde{g}}$ and $\Delta_{\tilde{w}}$ are the
contributions from the gluino and the wino loops to $b^{susy}_s$ and
$b^{susy}_2$ respectively and $M_U$ is the grand
unification scale. We notice that the quadratic Casimirs of
the SU(2), U(1) and SU(3) groups have the values 2,0 and 3 respectively.
Using a vector orthogonal to (2,0,3) and (1,1,1) we construct the
combination,
\begin{equation}
c= 2 \pi [ 3 \alpha^{-1}_2(M_Z)-\alpha^{-1}_{1Y}(M_Z)-2
\alpha^{-1}_s(M_Z)]. \label{c}
\end{equation}
Combining Eqn.(\ref{rge}) and Eqn.(\ref{c}) we have,
\bea
c=[3 b^{susy}_2-b^{susy}_1-2b^{susy}_s] \ln{M_U \over M_Z}
- 3 \Delta_{M_{\tilde{w}} } \ln {M_{\tilde{w}} \over M_Z} +2
\Delta_{\tilde{g}} \ln {M_{\tilde{g}} \over M_Z}. \label{comb}
\eea
In the right hand side of Eqn.(\ref{comb}) the gauge and the fermionic
contributions to the beta function coefficients cancel out due to the
orthogonality that has been mentioned earlier, where as the scalar
contribution does not. At this stage, using $\Delta_{\tilde{w}}=4/3$
and $\Delta_{\tilde{g}}=2$, we have the prediction of $\alpha_s$, as,
\be
\alpha^{-1}_s(M_Z)={ 1\over 2}~[3 \alpha^{-1}_2-\alpha^{-1}_1] -{3 \over
5 \pi} \ln{M_U \over M_Z} + { 1\over  \pi} \ln {M_{\tilde{w}} \over M_Z} -
{ 1\over  \pi} \ln {M_{\tilde{g}} \over M_Z}, \label{predic}
\ee
where, the second term is the threshold correction due to the doublet
triplet splitting, in which $M_U$ and $M_Z$ are the masses of the
superheahy triplet and the light doublet higgs scalars. It is interesting to
note that the `gluino'-term
and the `wino'-term in the right hand side of Eqn.(\ref{predic}) have a
relative sign among them.

\section{Non-universality of gaugino masses}

In the minimal scenario the spontaneous breaking of supergravity yields a
global supersymmetric theory supplemented by a set of soft
supersymmetry-breaking parameters. In particular in SUSY GUT
theories, gauge invariance implies that at the GUT scale, one must have
a universal gaugino mass ($m_{1/2}$). Using the RG evolution, if one runs
the gaugino masses from the GUT scale to the lower scales, a relation
between the mass of gluionos and the mass of the winos is gotten
\cite{nilles}. If
the masses of the s-particles are near the electroweak scale one gets,
\begin{equation}
x \equiv {m_{\tilde{g}}(M_Z) \over m_{\tilde{w}}(M_Z)} = {\alpha_{2}(M_Z)
\over \alpha_s (M_Z)} \sim  3.3. \label{ratio}
\end{equation}
This relation is of considerable significance regarding the prediction of
$\a_s$ in a supersymmetric GUT as we will explain now. To predict $\a_s$
we use the complete relation,
\begin{eqnarray}
\a^{-1}_s(M_Z)&=&{1 \over 2}~[3 \a^{-1}_2(M_Z) -\a^{-1}_{1Y}(M_Z)]
- {1 \over \pi} \ln{M_{\tilde{g}} \over M_Z}
{}~\theta(M_{\tilde{g}}-M_Z) \nonumber\\
&+& {1 \over \pi} \ln{M_{\tilde{w}} \over M_Z} ~\theta(M_{\tilde{w}}-M_Z)
 + T_{heavy} + T_{others}. \label{alstrong}
\end{eqnarray}
The theta functions in Eqn.(\ref{alstrong}) have the value 1
whenever the argument is positive and is zero otherwise. $T_{heavy}$
parametrizes the heavy threshold
corrections arising from the doublet-triplet splitting as well as from
the splitting among the heavy scalars. $T_{others}$ parametrizes the
effect of the light degrees of freedom apart from the winos and the
gluinos. We notice that, when the mass of the gluino is less than the mass
of Z,
neither wino nor the gluino contributes to Eqn(\ref{alstrong}). When the
mass of the gluino crosses the $M_Z$ range, the `gluino'-term, having a
negative
sign, starts contributing to the $\a^{-1}_s$, and consequently $\a_s$
starts to
increase. Only when the mass of the wino becomes comparable to the
electroweak scale, the `wino'-term starts contributing, which compensates
the increasing effect. In the Figure 1 \cite{mar} this phenomenon
is depicted \footnote{ There is a
10 \% decrease of $\a^{-1}_s$ at the 2 loop level.}. Notice that the
mass of the gluino is always more than the
mass of the wino due to the constraining relation given in Eqn.(\ref{ratio}).

\begin{figure}[htb]
\begin{center}
\epsfxsize=8.5cm
\epsfysize=8.5cm
\mbox{\hskip -1.0in}\epsfbox{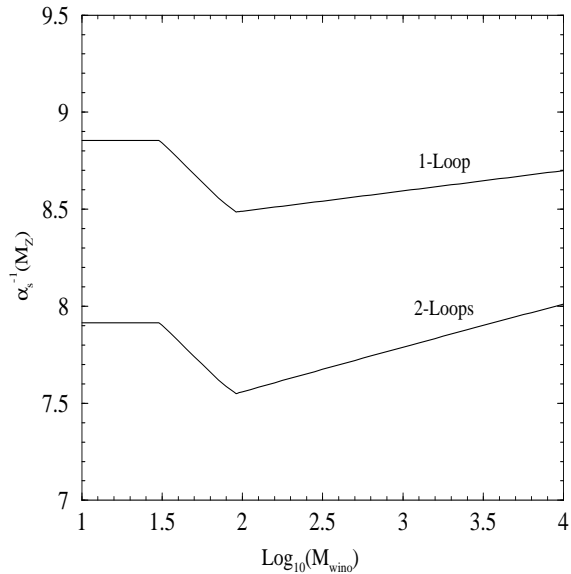}
\caption{ The one and two loop predictions of $\alpha_s(M_Z)$ in
MSSM with universal scalar mass term at the GUT scale.\label{}}
\end{center}
\end{figure}

Clearly, to reverse the effect, one needs to relax the mass relation
between the gluino and the wino \cite{nonuni}. This approach has been
taken by Roszkowski
and Shifman \cite{rs}. It has been noted by them that out of all soft
masses the soft masses of the wino and the gluino have the dominant
influence on the prediction of $\a_s$. The reason for this is two fold,

\noindent (a) The soft mass scale of the wino influences only the SU(2)
beta function coefficient, whereas the soft mass scale of the gluino
influences only the SU(3) beta function coeffcient only.

\noindent (b) The contribution of the wino and the gluino in the beta
function coefficients of SU(2) and SU(3) groups respectively are the
largest among all superparticles.

Along with this observations we may also observe that if the constraint in
Eqn.(\ref{ratio}) is
relaxed [the x parameter is varied], one can alter the prediction of
$\a_s$ considerably. The two loop predictions of $\a_s$ by choosing
different values of x have been summarized in Figure 2 \cite{rs}.

\begin{figure}[htb]
\begin{center}
\epsfxsize=8.5cm
\epsfysize=8.5cm
\mbox{\hskip -1.0in}\epsfbox{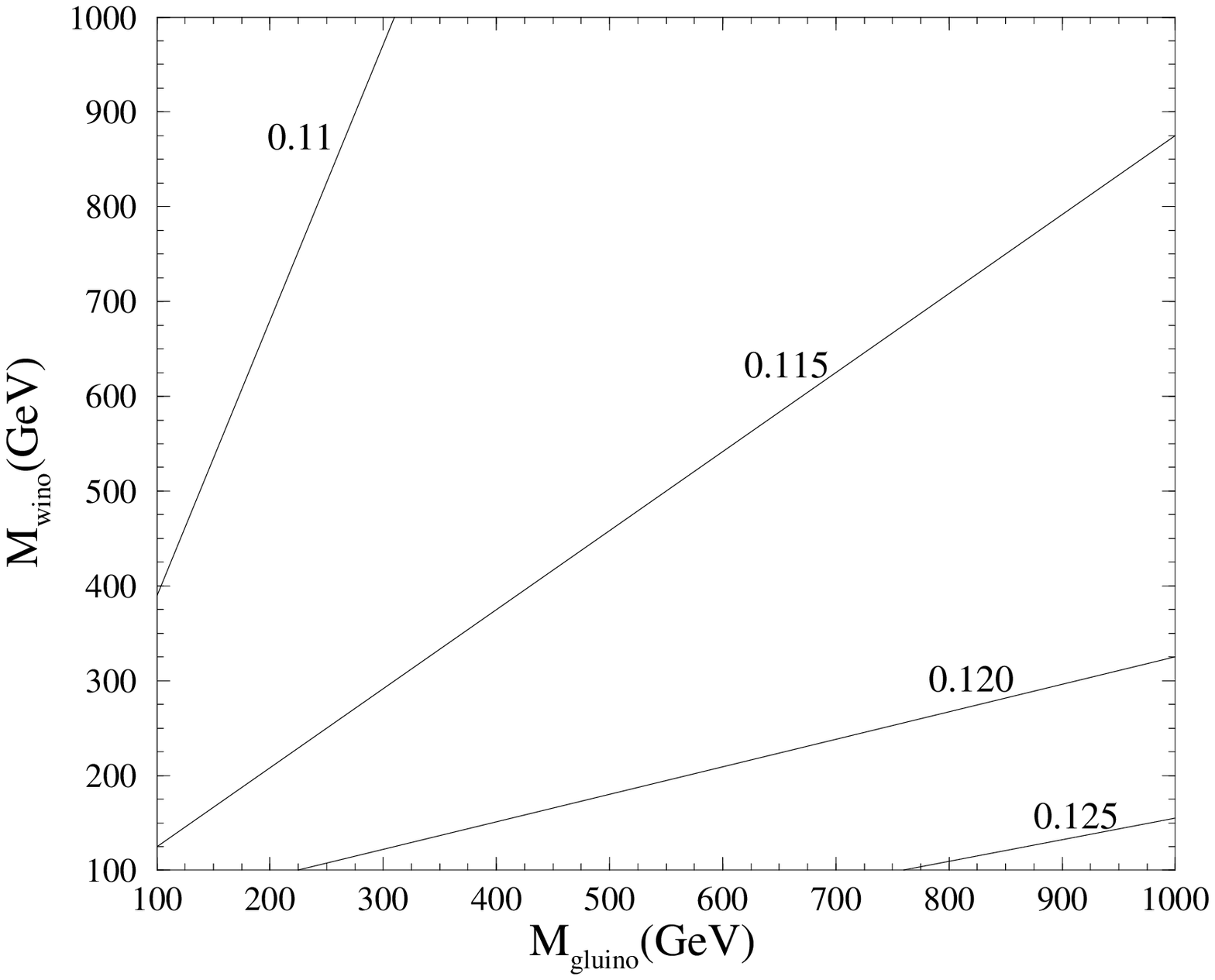}
\caption{ The contours of constant $\alpha_s (M_Z)$ for various
choices of the parameter `x'. \label{}} \end{center}
\end{figure}

One needs the gluino mass in the ball park of 100 GeV to predict
$\a_s=0.11$ by the above mechanism. Such a light gluino will have further
phenomenological consequences. For example, the gluino correction enhances
the hadronic width of $Z$. However, the $Z \rightarrow b
\overline{b}$ width increases too much for $M_{\tilde{g}} \sim 100$ GeV,
and one has to descend to unacceptably low sqark masses to reconcile with
the experimental $Z \rightarrow b \overline{b}$ width.

\section{Heavy thresholds and missing doublet SU(5)}

In the presence of heavy threshold corrections, the unification scale is
no longer well-defined. Let us, therefore, define a scale $\Lambda$,
which is larger than any GUT scale mass. Above the scale $\Lambda$ all the
couplings remain unified. Again we will consider a toy SU(5) example. Now,
as we are interested in the heavy thresholds only, let all the
superpartners of the standard model fermions and gauge bosons be
degenerate at the scale $M_Z$. The heavy spectrum of the minimal SU(5)
model is simple. The $(3, 2, 5/6) + (\overline{3}, 2, 5/6)$ components of
the $24$-scalar get absorbed by the heavy gauge bosons with a common mass
$M_V$. The SU(3)-octet and the SU(2)-triplet have a common mass
$M_\Sigma$ whereas the singlet has a mass $0.2M_\Sigma$. We can relate
the three gauge couplings at the scale $M_Z$ to the unified coupling at
the scale $\Lambda$ as \cite{hisano},
\begin{eqnarray}
\alpha_s^{-1}(M_Z)&=&[\alpha_U^{-1}(\Lambda) - { 2 \over \pi} \ln{
\Lambda \over M_V} + { 3 \over 2 \pi} \ln{ \Lambda \over M_\Sigma} +
{1 \over 2 \pi} \ln { \Lambda \over M_{H_3}}]
+ { 1 \over 2 \pi} b_s^{susy} \ln{\Lambda \over M_Z}, \nonumber\\
\alpha_2^{-1}(M_Z)&=&[\alpha_U^{-1}(\Lambda) - { 3 \over \pi}
\ln{\Lambda \over M_V} + {1 \over \pi} \ln{ \Lambda \over M_\Sigma}]
+ { 1 \over 2 \pi} b_2^{susy} \ln{\Lambda \over M_Z},
\nonumber\\
\alpha_1^{-1}(M_Z)&=&[\alpha_U^{-1}(\Lambda) - { 5 \over \pi}
\ln{\Lambda \over M_V} + {1 \over 5 \pi} \ln { \Lambda \over
M_{H_3}}]+{1 \over 2 \pi} b_1^{susy} \ln{\Lambda \over M_Z}.
\end{eqnarray}
Taking the combination c as in the previous section,
 we find that it is independent of any field that
is in the adjoint of SU(5) and we recover the result,
\begin{equation}
\alpha_s^{-1}(M_Z)={1 \over 2} [3 \alpha_2^{-1}-\alpha_1^{-1}] - { 3 \over
5 \pi} \ln{M_{H_3} \over M_Z}.
\end{equation}
Now, it is clear why we interpreted the second term in Eqn. (\ref{predic}) as
the effect of doublet triplet splitting.

We notice a special property of the $24$-dimensional scalar in relation
to the combination c. While the components $(3,2,5/6) + (\overline{3},2,
-5/6)$ are absorbed as the logitudinal components of the heavy SU(5)
gauge bosons, the rest are in the adjoint representation of the low
energy groups. We have already noticed that fields which are in the
adjoint representation of the low energy groups and having a common mass
cannot contribute to the combination c. Now, we also note that even the
would be goldstone bosons do not contribute to c; and no other component
of the adjoint is left; except the singlet which does not contribute to
the beta function coefficients. So the splitting in the adjoint $24$,
does not affect the prediction of $\alpha_s$ at all.

The situation changes if we introduce 75-dimensional scalar instead of
the adjoint to break the unification symmetry along with the $50$ and
$\overline{50}$ dimensional representations needed for the missing
doublet mechanism \cite{masiero}. Twelve components of $75$ will be eaten
up by the
the heavy gauge bosons. Remaining 63 components will be split in mass
\cite{yamada}.
Moreover, the color triplets from $50$ and $\overline{50}$ Higgs scalars
get mixed with the color triplets of $5$ and the $\overline{5}$. The rest
of the components of $50$ and the $\overline{50}$ components are not
split among themselves. They acquire a common mass $M_{50}$. The
typical masses of the various GUT scale heavy scalars \cite{yamada} have
been summerized in Table\ref{split}.

\begin{table}[htb]
\begin{center}
\[
\begin{array}{|c||c||c|}
\hline
component&mass &comments\\
\hline
(8,3,0) & M_\Sigma &   \\
(3,1, \pm 5/3)  &  0.8 M_\Sigma &   \\
(6,2, \pm 5/6) &  0.4 M_\Sigma &in~75   \\
(1,1,0)        &  0.4 M_\Sigma &of~SU(5)   \\
(8,1,0)        &  0.2 M_\Sigma &   \\
\hline
(3,1, \pm 1/3) &  M_{D_1}& in~5+\overline{5}  \\
(3,1, \pm 1/3) &  M_{D_2}& and~50+\overline{50}  \\
\hline
rest ~of~ 50+\overline{50}&M_{50} &   \\
\hline
\end{array}
\]
\end{center}
\caption{The representations and masses of the heavy scalars in the
missing doublet SU(5) model. Bars have been suppressed in certain
representation for the compactness of presentation}
\label{split}
\end{table}

Given the masses and the transformation properties in Table\ref{split},
it is straightforward to calculate the change in the prediction of
$\alpha_s(M_Z)$. Indeed such a calculation have been performed in
Ref.\cite{bagger} in which they have carefully taken into account both
the heavy and the light threshold effects. They conclude that a large
negative correction to $\alpha_s$ comes in the missing doublet model due
to the splitting in the $75$ Higgs. The value of $\alpha_s=0.112$ is
achievable (taking into account a lower bound on a combination of
$M_{D_1}$ and $M_{D_2}$ from proton decay) with the s-particle masses
around the electroweak scale. For further details the reader is referred
to their papers.

\section{Intermediate scale in supersymmetric SO(10)}

There is a general understanding in the literature that the LEP measurements
of the gauge couplings at the scale $M_Z$ is a hint to a one step unified
theory. There are however several physical arguments based on neutrino
physics \cite{hdm,numass} as well as strong CP problem \cite{kim} suggesting
that there may
be an intermediate scale corresponding to a gauged (local) B-L symmetry
breaking somewhere around $10^{11}$ to $10^{12}$ GeV. However, such a scale
must not affect the gauge unification constraints. To examine this, let
us write down the RGE of the three couplings
relating them to the gauge coupling at
the unification scale $M_U$ to the ones at
the weak scale $M_Z$ and introducing a general intermediate scale $M_I$, as,
\begin{eqnarray}
2 \pi \a_1^{-1}(M_Z)&=& 2 \pi \a_U^{-1}(M_U) + b_1 \ln{M_I \over M_Z} + b_1
^\prime {M_U \over M_I}, \nonumber\\
2 \pi \a_2^{-1}(M_Z)&=& 2 \pi \a_U^{-1}(M_U) + b_2 \ln{M_I \over M_Z} +
b_2^\prime {M_U \over M_I}, \nonumber\\
2 \pi \a_s^{-1}(M_Z)&=& 2 \pi \a_U^{-1}(M_U) + b_s \ln{M_I \over M_Z}
+b_s^\prime {M_U \over M_I}. \label{rgint}
\end{eqnarray}
At first, we notice that the one-loop beta function coefficients $b_i$ in
MSSM
for the groups SU(3), SU(2) and U(1) are -3, 1 and 33/5 respectively.
Presently we are interested in the beta function coefficients
$b_i^\prime$ governing the slopes in the region between $M_I$ and $M_U$.
So, by taking a vector orthogonal to \\(-3,1,33/5) and {(1,1,1)} we
construct a second combination $c_1$ in which $b_i$'s get eliminated but
$b_i^\prime$ s survive, namely,
\be
c_1=2 \pi (7 \a_s^{-1} - 12 \a_2^{-1} + 5 \a_1^{-1}). \label{c2}
\ee
We combine, Eqn.(\ref{rgint}) and Eqn.(\ref{c2}) to get,
\be
c_1=(7 b_s -12 b_2 + 5 b_1) \ln {M_I \over M_Z} +
(7 b_s^{\prime} -12 b_2^\prime +5 b_1^\prime) \ln {M_U \over M_I}
\label{c2cond}.
\ee
Due to the orthogonality in the construction of
$c_1$ the coefficient of $\ln (M_I / M_Z)$
vanishes and we are left with,
\be
c_1=(7 b_s^\prime -12 b_2^\prime +5 b_1^\prime)
\ln {M_U \over M_I} \label{c2b}.
\ee
Moreover, if we assume that the values of $\a_1$, $\a_2$ and
$\a_s$ are such that an one-step unification is possible with
MSSM and a big desert - we get a condition from Eqn.(\ref{c2cond}) in the
limit $M_U=M_I$, which is,
\be
c_1=0. \label{c20}
\ee
Combining, Eqn.(\ref{c20}) and Eqn.(\ref{c2b}), we get,
\be
(7 b_s^\prime -12 b_2^\prime +5 b_1^\prime)
\ln {M_U \over M_I} =0
\label{c2c}.
\ee
We note the two solutions of Eqn.(\ref{c2c}).

\noindent (a) $M_U=M_I$ which leads us back to the one-step unification.

\noindent (b) Due to the presence of new fields above the scale $M_I$, the
beta function coefficients $b_i^\prime$ conspire among themselves to
produce\cite{sato},
\be
7 b_s^\prime -12 b_2^\prime +5 b_1^\prime=0.
\label{cond}
\ee
One can restrict the type of Higgs representations above the intermediate
scale $M_I$ by requiring that the supersymmetric SO(10) GUT emerges from an
underlying
superstring theory. If we restrict ourselves to only those Higgs scalars
which can arise from simple superstring models with Kac-Moody levels one
or two, we can have a restricted number of solutions of Eqn.(\ref{cond}).
We will consider an intermediate symmetry group ${G_I=U_{B-L}\times SU(2)_L
\times SU(2)_R \times SU(3)_C}$. The solutions can be characterized by three
integers ($n_C,n_H,n_X$), where $n_C$ refers to the number of
(0,1,1,8), $n_H$ means the number of (0,2,2,1)
fields and $n_X$ means the number of {(1,1,2,1)+(-1,1,2,1)} fields under
the intermediate symmetry gauge group $G_I$. The various
scenarios\cite{leemoh,we} are tabulated in Table\ref{table2}.
\begin{table}[htb]
\begin{center}
\[
\begin{array}{|c||c||c||c||c||c||c||c|}
\hline
model&I &II & III& IV&V&VI&VII\\
\hline
(n_C,n_H,n_X)&(0,1,2)&(0,1,3) &(0,1,4) &(0,2,3) &(0,2,4)&(0,2,5)&(1,2,4)  \\
\hline
\end{array}
\]
\end{center}
\caption{The various models which satisfy the condition in Eqn.~15
approximately.} \label{table2}
\end{table}

\begin{figure}[htb]
\begin{center}
\epsfxsize=8.5cm
\epsfysize=8.5cm
\mbox{\hskip -1.0in}\epsfbox{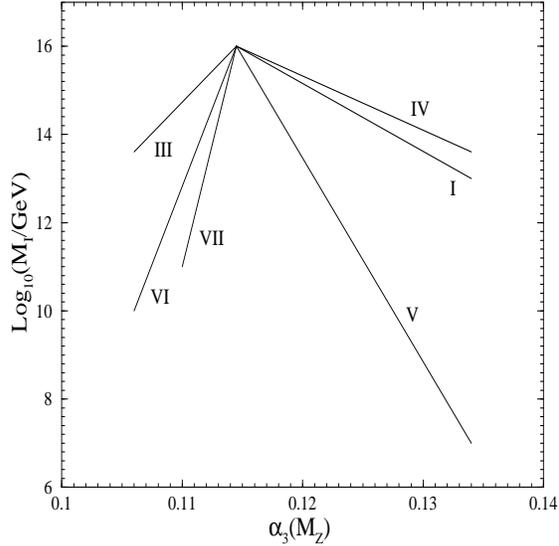}
\caption{ The predictions of $\a_s$ are displayed as constant $\a_s$
contours for the various models given in Table 2.\label{fig3}}
\end{center}
\end{figure}

The predictions of $\a_s$ for different models \cite{leemoh,we} have been
plotted in Figure \ref{fig3}. We notice from Figure \ref{fig3} that the
models VI and VII
are particularly interesting for the value of $\alpha_s$ in the range of
0.11. A comment is in order. The
value of the intermediate scale ($M_I$) is a new parameter. But we notice
that even this new parameter is quite constrained by the the LEP measurements
of the couplings at the scale $M_Z$. Moreover, only three out of the seven
intermediate scale models in Table\ref{table2} will survive if
the value of $\a_s(M_Z)$ turns to be 0.112.

There is another interesting aspect of such intermediate scale SO(10)
GUTs. We see that the requirement of unification forces us to introduce
new scalar fields above the scale of $B-L$ symmetry breaking. These new
scalars will also affect the evolution of Yukawa couplings above the
intermediate symmetry breaking scale. The evolution of various Yukawa
couplings have been calculated in Ref \cite{we}. The results are
tabulated in Table\ref{table3} and Table\ref{table4}.

\begin{table}[htb]
\begin{center}
\[
\begin{array}{|c||c||c||c||c||c||c||c||c|}
\hline
Y_1&Y_2&h_t(M_t)&h_b(M_t)&h_\tau(M_t)&
tan\beta&m_b(M_t)&m_t(M_t)&{m_b(M_t) \over m_\tau(M_t)}\\
\hline
1&1         &1.010&0.96&0.62&60.43&2.77&176.83& 1.56 \\
1&10^{-1}   &1.060&0.84&0.52&51.26&2.85&184.46&1.60  \\
1&10^{-2}   &1.094&0.460&0.270&26.7&3.01&190.34&1.69  \\
1&10^{-3}   &1.103&0.160&0.095&9.25&3.06&190.90&1.72  \\
1&10^{-4}   &1.104&0.054&0.030&2.80&3.07&181.03&1.73  \\
1&{1 \over 2}10^{-4}   &1.104&0.037&0.021&1.85&3.08&169.16&1.73 \\
\hline
\end{array}
\]
\end{center}
\caption{The values of $h_t(M_t)$, $h_b(M_t)$, $h_\tau(M_t)$ and
calculated by RGE for
$\alpha_s=0.11$ in model VI. The prediction of the
masses $m_b$ and $m_t$ at the scale
$M_t$ has been quoted in GeV. $M_t$ is defined as 170 GeV. $tan \beta$
has been calculated assuming $m_\tau(M_Z)=1.777$ GeV.} \label{table3}
\end{table}

\begin{table}[htb]
\begin{center}
\[
\begin{array}{|c||c||c||c||c||c||c||c||c|}
\hline
Y_1&Y_2&h_t(M_t)&h_b(M_t)&h_\tau(M_t)&
tan\beta&m_b(M_t)&m_t(M_t)&{m_b(M_t) \over m_\tau(M_t)}\\
\hline
1&1         &0.99&0.960&0.59 &57.81   &2.87&173.90&1.61 \\
1&10^{-1}   &1.05&0.830&0.50 &48.78   &2.97&182.65&1.67  \\
1&10^{-2}   &1.09&0.460&0.26 &25.41   &3.17&189.11&1.78  \\
1&10^{-3}   &1.10&0.160&0.09 &8.81    &3.23&189.69&1.82  \\
1&10^{-4}   &1.10&0.052&0.023 &2.65   &3.24&178.80&1.82  \\
1&{1 \over 2}10^{-4}   &1.104&0.037&0.020&1.74&3.24&165.70&1.82 \\
\hline
\end{array}
\]
\end{center}
\caption{The values of $h_t(M_t)$, $h_b(M_t)$, $h_\tau(M_t)$ and
calculated by RGE for $\alpha_s=0.11$ in model VII. The
prediction of the masses $m_b$ and $m_t$
at the scale $M_t$ has been quoted in GeV. $M_t$ is
defined as 170 GeV. $tan \beta$ has been calculated assuming
$m_\tau(M_Z)=1.777$ GeV.} \label{table4}
\end{table}

It is widely known that in SUSY GUTs with one step breaking predict a
large value of $m_b$ for the major part of the parameter space \cite{barg}.
The study of b-$\tau$ unification including a right handed neutrino has also
been performed \cite{fv}. However, in this study no new
gauge interactions beyond the intermediate scale was considered
and due to renormalization effects of the new Yukawa coupling,
a 10-15\% increase in the mass of the b-quark was obtained. However one
sees that the inclusion of the new left-right symmetric gauge and Higgs
interactions at $M_{B-L}$ surviving from a string inspired  SO(10) GUT, and
constrained by Eqn.(\ref{cond}), the running of the b-quark Yukawa coupling
is altered and as a result an attractive reconciliation with the
experimental measurements \footnote{ Taking $m_b(m_b)$ in the range
4.1 to 4.5 GeV and $m_\tau(m_\tau)$ to be 1.777 GeV \cite{pdg}, one
typically gets ${m_b(M_t) \over m_\tau(M_t)}$ in the range 1.7 to 1.9 \cite{
barg, fv}.} can be achieved.

\section{Mini doublet-triplet splitting}

In this section we will explore a possible reverse doublet-triplet
splitting \cite{marbr}
which will have an effect opposite to the conventional doublet-triplet
splitting on the prediction of $\alpha_s$.  Such a strange reverse
doublet-triplet splitting is indeed possible in a SO(10) model when
there is a mechanism to strongly suppress the Higgsino mediated proton
decay \cite{babr} as will be displayed below.

We consider the prediction of $\alpha_s$
including the threshold effects in SUSY SU(5), which is well-studied
in the literature \cite{barhall,yamada,hisano,mar,bagger}. Throughout
this section we will assume that including the threshold corrections,
the minimal SUSY SU(5) GUT predicts $\alpha_s=0.126$ \cite{bagger}; we
will also assume that the mass of the color triplet Higgs scalars in a
minimal SU(5) GUT is $10^{16.6}$ GeV \cite{mar}. In particular the
prediction of $ \alpha_s$ in the minimal SUSY SU(5) can be written as,

\begin{equation}
\alpha^{-1}_s(M_Z)={1 \over 2} ~[3 \alpha^{-1}_2(M_Z)-\alpha^{-1}_1(M_Z)]
- { 3 \over 5 \pi} \ln[{M_3 \over M_2}] + T_L \label{a3su5},
\end{equation}

Where, $M_3$ and $M_2$ are the masses of the triplet and the doublet
Higgs scalars present in the $5$ and $\overline{5}$ representations of
SU(5), and $T_L$ parametrizes the contribution from all other light
degrees of freedom (excluding the light Higgs doublets) \cite{hisano},
and in a simple step function approximation\footnote{ $M_{SUSY}$ can
be considered in the simplest approach as a common susy breaking
scale, or as an effective susy mass parameter \cite{lang} resuming the
effect of the detailed susy spectrum, and in this sense it can be
either more or less than $M_Z$ depending on the super-partner masses.}
$T_L= {1 \over 2 \pi}
\ln {M_{SUSY} \over M_Z}$. In the minimal model the triplet-mass,
which is bounded from below from the non-observation of proton decay,
remains at the GUT scale. On the contrary the mass of the doublet is
of the order of the electroweak scale. In such a generic situation,
that is, whenever $M_3>M_2$ the doublet-triplet splitting increases
the prediction of $\alpha_s$. However, notice the hypothetical
possibility, that if the mass of the doublet were more than the mass
of the triplet, we would have had a reverse effect on $\alpha_s$.
Keeping this in mind we add one more $5+\overline{5}$ Higgs scalars
with doublet and triplet masses as $M^\prime_2$ and $M^\prime_3$ GeV
respectively. In that case the Eqn.(\ref{a3su5}) gets modified to,
\begin{equation}
{\alpha^\prime}^{-1}_s(M_Z)={1 \over 2} ~[3
\alpha^{-1}_2(M_Z)-\alpha^{-1}_1(M_Z)] - { 3 \over 5
\pi}
\ln[{M_3 M^\prime_3
\over M_2 M^\prime_2}] + T_L \label{a3}.
\end{equation}
Taking the difference of Eqn.(\ref{a3su5}) and Eqn.(\ref{a3}) and
assuming,
\begin{equation}
M_3=10^{16.6}~;~M_2=10^2~;~M^\prime_3=10^x~;~M^\prime_2=10^y,
\end{equation}
we get,
\begin{equation}
\Delta \alpha^{-1}_s={\alpha^\prime}^{-1}_s(M_Z)-\alpha^{-1}_s(M_Z)
={3 \over 5 \pi} (y-x)
\ln 10. \label{diff}
\end{equation}
It is easy to check from Eqn.(\ref{diff}) that taking $y-x=2.26$ we
can get $\Delta \alpha^{-1}_s =0.99$ and consequently $\alpha_s$
decreases by 11\%, from 0.126 to 0.112.  Instead if we add n extra
pairs of $5+\overline{5}$ the required splitting in each SU(5)
multiplet is only 2.26/n orders of magnitude.

We can give an example of incorporating this mechanism in a realistic
SUSY SO(10) GUT. The main problem to lower the mass of the color
triplet Higgs Scalars comes out of the stringent experimental upper
bounds imposed on the amplitude of the Higgsino mediated proton decay
diagrams. Babu and Barr \cite{babr} have shown that it is possible to
suppress the Higgsino mediated proton decay strongly in an SO(10)
model by a judicious choice of the fields, couplings and VEVs at the
GUT scale. Consider the SO(10) invariant superpotential
\cite{babr}
\begin{equation}
W= \lambda 10_{1H} 45_H 10_{2H}+\lambda^\prime 10_{2H} 45^\prime_H
10_{3H}+M 10_{3H} 10_{3H}+ \sum^{3}_{i,j=1} f_{ij} 16_i 16_j 10_{1H}.
\end{equation}
If $45$ and $45^\prime$ get VEVs in the directions \cite{dw,babr}
\begin{equation}
\langle 45\rangle =  \eta \otimes diag(a,a,a,0,0)~~;~~
\langle 45^\prime \rangle  =  \eta \otimes diag(0,0,0,b,b)~~;~~
\eta  \equiv  \pmatrix{0&1 \cr -1&0},
\end{equation}
the super-heavy mass matrices of the doublets and the triplets are of
the form,
\begin{equation}
\pmatrix{\overline{2}_1& \overline{2}_2 & \overline{2}_3}
\pmatrix{0&0&0\cr0&0&\lambda^\prime b\cr0&-\lambda^\prime b &M}
\pmatrix{2_1 \cr 2_2 \cr 2_3}
{}~~{\rm and}~~
\pmatrix{\overline{3}_1& \overline{3}_2 & \overline{3}_3}
\pmatrix{0& \lambda a&0\cr -\lambda a &0& 0 \cr 0& 0 &M }
\pmatrix{3_1 \cr 3_2 \cr 3_3}.
\end{equation}
The absence of any direct coupling between $\overline{3}_1$ and $3_1$
suppresses the Higgsino mediated proton decay. The absolute values of
the masses for the doublets ($M_D$) and the triplets ($M_T$) are given
by\footnote{ When supersymmetry is broken the masses receive
correction of the order of $m_{3/2}$.}, \begin{equation} M_D=(0,{M \pm
\sqrt{M^2-{\lambda^\prime}^2b^2} \over 2})~~;~~ M_T=(\lambda a,\lambda
a,M).
\end{equation}
There can be various choices of parameter space leading to the
required lowering in the prediction of $\alpha_s$ [see
Eqn.(\ref{diff})]. The simplest choice is, $ M^2={\lambda^\prime}^2
b^2.  $ In this case the masses are,
\begin{equation}
M_D=(0,M/2,M/2) ~~{\rm and}~~ M_T=(\lambda a, \lambda a, M).
\end{equation}
We notice that, in this model one pair of doublet-triplet is almost
degenerate.  Now, using the fact that in minimal SUSY SU(5) the mass
of the color triplet is $10^{16.6}$ GeV and using,
\begin{equation}
\lambda a = 10^x~~,~~M=10^y,
\end{equation}
we get,
\begin{equation}
\Delta \alpha^{-1}_s={3 \over 5 \pi}[(16.6-x)+(y-x)-{\ln 4 \over \ln 10}]
\ln 10 . \label{dsso10}
\end{equation}
We can achieve the desired suppression of $\alpha_s$ if for example
$x=15.2$ and $y=16.6$. We expect that such a splitting is not
difficult to achieve given the number of parameters in the SO(10)
invariant superpotential. The threshold effects due to heavy SO(10) gauge
bosons have been discussed in Ref \cite{marbr}.

\section{Conclusion}

To summarize, we have discussed various ways to reconcile a possible
low value of $\a_s$ with the idea that all gauge couplings eventually
unify . They all indicate new physics beyond the canonical minimal
single SUSY GUT scenarios and point towards models with intermediate
scales\cite{we,martin} or new Higgs fields at the GUT scale\cite{marbr}
or perhaps even non-GUT type physics\cite{rs}. Any realistic model
building in the SUSY GUT framework must therefore respect either one
of these scenarios. There are of course other suggestions to change
the predictions of $\a_s$ by including gravitational effects at the
GUT scale\cite{utpal,nath} provided these effects are assumed to occur
with an enhanced strength, a possibility which may not be so natural
in many theories (although one cannot be completely sure about
the strength of the gravitational effects at this stage of our knowledge).

\section{Acknowledgements}

B.B would like to thank Mar Bastero-Gil for a number of discussions and
computational help.

\end{document}